\documentclass{aastex63}

\usepackage{graphicx}
\usepackage{epsfig}
\usepackage{times}
\usepackage{natbib}
\usepackage{url}
\usepackage{color}
\usepackage{amssymb,amsmath}
\usepackage{hyperref}

\shorttitle{Speed inconsistency between EUV waves and type IIs}

\begin{document}

\title{Global Nature of Solar Coronal Shock Waves shown by Inconsistency between EUV Waves and Type II Radio Bursts}

\correspondingauthor{Ryun-Young Kwon}
\email{rkwon@kasi.re.kr}

\author[0000-0003-2973-5847]{Aarti Fulara}
\affil{Aryabhatta Research Institute of Observational Sciences, Manora Peak, Nainital, 263001, India}

\author[0000-0002-2106-9168]{Ryun-Young Kwon}
\affil{Korea Astronomy and Space Science Institute, Daejeon 34055, Republic of Korea}

\begin{abstract}
We re-examine the physical relationship between Extreme-UltraViolet (EUV) waves and type II radio bursts. It has been often thought that they are two observational aspects of a single coronal shock wave. However, a lack of their speed correlation hampers the understanding of their respective (or common) natures in a single phenomenon. Knowing the uncertainties in identifying true wave components from observations and measuring their speeds, we re-examine the speeds of EUV waves reported in previous literature and compare these with type II radio bursts and Coronal Mass Ejections (CMEs). This confirms the inconsistency between the speeds of EUV waves and their associated type II radio bursts. Second, CME speeds are found to have a better correlation with type II radio bursts than EUV waves. Finally, there exists a tendency for type II speeds and their range to be much greater than those of EUV waves. We demonstrate that the speed inconsistency is in fact an intrinsic tendency and elucidate the nature of a coronal shock wave consisting of both driven and non--driven parts. This suggests that the speed inconsistency would remain even if all other uncertainties were removed.
\end{abstract}
\keywords{Solar coronal waves(1995) --- Solar coronal radio emission(1993) --- Solar coronal mass ejections(310)}

\section{Introduction} \label{S-Intro}
Solar eruptive events, such as flares and Coronal Mass Ejections (CMEs), release huge amounts of energy. The profound effects of such explosive processes are not confined to their parent active regions, but also causes disturbances over a wide spatial range in the forms of fast magnetosonic waves and shocks \citep[e.g.,][]{Cliver95,Kwon13}. The existence of these coronal waves has long been established based on metric type II radio bursts \citep[e.g.,][ hereafter type IIs]{Wild50, Cliver99}. Such disturbances have also been observed in optical observation as arc-shaped bright fronts of H$\alpha$ wings, indicative of propagating disturbances in the chromospheric layer \citep{Moreton60,MR60}; such phenomena are termed Moreton or Moreton–Ramsey waves. Moreton–Ramsey waves were initially interpreted as the chromospheric response to the fast magnetosonic wave traveling in the solar corona \citep{Uchida1968}. Thus, they were thought to be remote-sensing observations of the shocks responsible for type IIs \citep{Uchida1974}. When arc-shaped bright fronts were also observed with Extreme Ultraviolet Imaging Telescope \citep[EIT;][]{Del95} onboard Solar and Heliospheric Observatory \citep[SOHO;][]{Domingo95} spacecraft, they were immediately interpreted as the {\it coronal} waves that are the origin of the {\it chromospheric} Moreton--Ramsey waves \citep{Moses97,Thompson98}. They were first named for the instrument that discovered this phenomenon, i.e., EIT waves, but later came to be called EUV waves since they are generally identified in EUV observations. EUV waves have a broad speed distribution, from several tens of km s$^{-1}$to more than 1000 km s$^{-1}$ \citep[see review by][]{Warmuth15, Long2017a}. 

Due to the lack of coronagraphic observations in the 1960s and 1970s, the agent of the coronal shocks observed as type IIs and Moreton--Ramsey waves was thought to be flares; however, since 1995, the regular space-based coronagraphic observations of SOHO LASCO \citep{Brueckner95} have established that they have a stronger association with CMEs than flares. \citet{Gopalswamy05, Gopal09} have shown that metric type II emission is usually driven by CMEs. Support for the CME-driven shock scenario is given by observations of broadening and intensity changes in the UV emission lines ahead of the CME front, attributed to shocks associated with type IIs \citep{Mancuso02, Cia05}. Similarly, analyzing 173 EUV waves observed between 1997 and 1998 by SOHO EIT, \citet{Bie02} found an {\it unambiguous} relationship between EUV waves and CMEs, while EUV waves with {\it bright} and {\it sharp} fronts also have a strong relationship with flares.

It has been often thought that EUV waves and type IIs are two observational aspects of a single coronal shock wave, but physical inconsistencies between them hampers understanding of their different or common nature. The inconsistency is two-fold based on their one-to-one correspondence and speed. \citet{Muhr14} analyzed 60 strong EUV wave events from January 2007 to February 2011 and found a 22\% association of EUV waves with type IIs. \citet{Nitta13} presented a statistical analysis of 138 events, finding a 54\% association. \citet{Nitta14} presented examples in which a type II is not associated with an EUV wave and {\it vice versa}. They suggested that neither EUV waves nor type II bursts serve as a necessary condition for coronal shock waves. More recently, \citet{Long17} found that  out of 164 events, 40\% of the EUV waves are associated with type IIs.

Few studies have compared their speeds between EUV waves and type IIs. \citet{kl00} found that they are strongly associated (90\%), but their speeds remain poorly correlated. Recently, \citet{Long17} found that the median speeds for individual EUV wave event correlate closely with their accelerations, indicating that the median speed is a physically meaningful characteristic. We interpret that the random error in speed has less effect on the median values. While systematic errors may not be removed by taking the median value, the correlation would have less of an effect on the systematic error. Interestingly, however, they found that even their median speeds are not correlated with the speeds of type IIs. Such inconsistency was also found by \citet{Warmuth10}.

Supposing that EUV waves are the imaging observations of fast magnetosonic shock waves, which are themselves observed as type IIs in radio observations, several possible explanations may account for this inconsistency. First, speed inconsistencies could result from different propagating directions. As indicated by their slow frequency drift, type IIs are likely to be a shock driven by the CME radial motion. On the other hand, it is widely accepted that EUV waves are driven by the lateral expansion of their associated CME. Thus, the directionality naturally explains their different speeds. However, it is also well-known that CME radial speeds and lateral expansion speeds are strongly correlated \citep{Dal03, Schwenn05}, suggesting that the directionality itself should not affect their correlations. For instance, \citet{Warmuth10} showed that the speeds of Moreton--Ramsey waves closely correlate with those of type IIs. This has provided the basis for various non-wave interpretations of EUV waves \citep[see][for reviews]{Warmuth15, Long2017a}; in contrast, Moreton--Ramsey waves have been widely accepted to be a type II-related wave phenomenon.

Secondly, it may result from uncertainties inherent to measuring their speeds. In case the same fronts are given, the speeds measured by various people with different methods will strongly correlate one another, and thus these differences in the speeds will not significantly affect the correlation when comparing them with other parameters. Because the fronts to track for type IIs and CME noses are clearly given, the largest uncertainties in correlations should arise from the fact that the EUV observations show not only the wave fronts but also non-wave components \citep{Warmuth15, Long2017a}. Taking the occurrence of slower non-wave components following a true wave front into consideration \citep[e.g.,][]{Chen11}, one may consider only the faster one when determining the wave speed. However, EUV waves are also subject to the projection effect, and thus the projected fronts could appear faster than the actual ones \citep{Kwon13, Lario14, Downs21}. Such complicated appearance in an event inhibits the measurements of their accurate speeds and errors. Alternatively, we analyze EUV waves listed in \citet{Nitta13}, in which the speeds are used for statistical studies \citep{Nitta13,Nitta14,Long17}. We take advantage of these previous speed measurements to find out if there exists a physically meaningful trend of tendency between EUV waves and type IIs.

In this paper, we re-examine the physical relationship between EUV waves and type IIs. We have identified 60 EUV wave events associated with type IIs from a catalog compiled by \citet{Nitta13}. In Section \ref{sec:res}, we show comparisons among these EUV waves, type IIs, and parent CMEs to evaluate whether the inconsistency is due to speeds uncertainties. The speeds determined by \citet{Nitta13} and \citet{Long17} are also presented for the comparisons. Finally, in Section \ref{sec:DC}, we present discussions about our findings and conclusions.

\section{Results} \label{sec:res}
Since errors in speeds of type IIs and CMEs would have small effects on the correlations as discussed in Section \ref{S-Intro}, we used the speeds previously reported in the LASCO CDAW catalog \citep{y04} and the NOAA list for CMEs and type IIs, respectively.
Knowing the uncertainties described in Section \ref{S-Intro} and thus to be compared with previous measurements, we analyzed the EUV wave events listed in \citet{Nitta13} which are already extensively studied \citep{Nitta13,Nitta14,Long17}. We used the Solar Dynamics Observatory \citep[SDO;][]{Pesnell12}/ Atmospheric Imaging Assembly \citep[AIA;][]{Lemen12} AIA 193 \AA\ images to re-examine the properties of the EUV waves. We compared the timing of these events with CMEs and type IIs in the LASCO CDAW catalog and the NOAA list, respectively. As a result, we found 60 EUV wave events associated with type IIs. The list of the events investigated in this study and the physical parameters are given in Table 1.

\begin{figure*}
\gridline{\fig{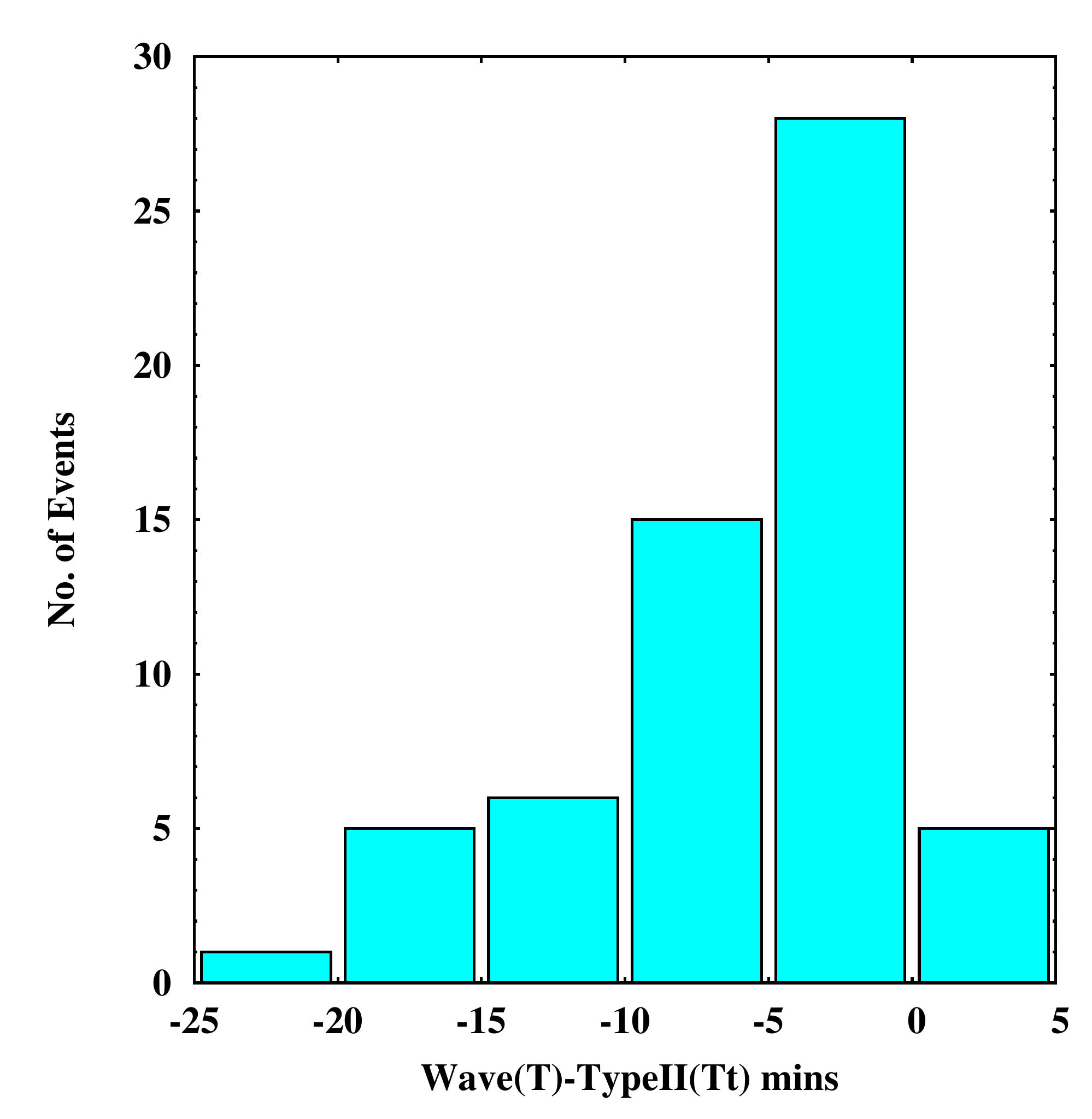}{0.33\textwidth}{(a)}
          \fig{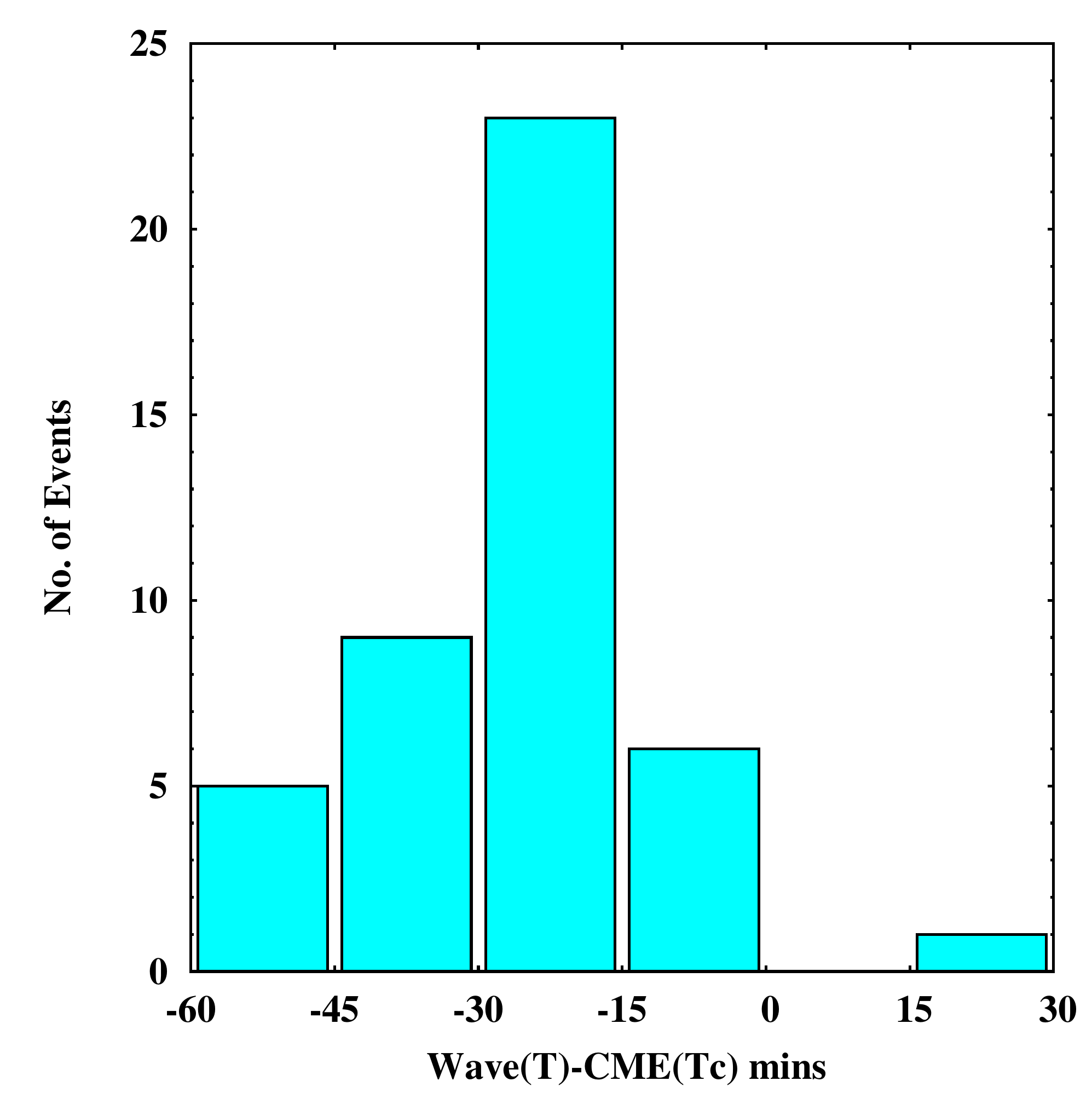}{0.33\textwidth}{(b)}
          \fig{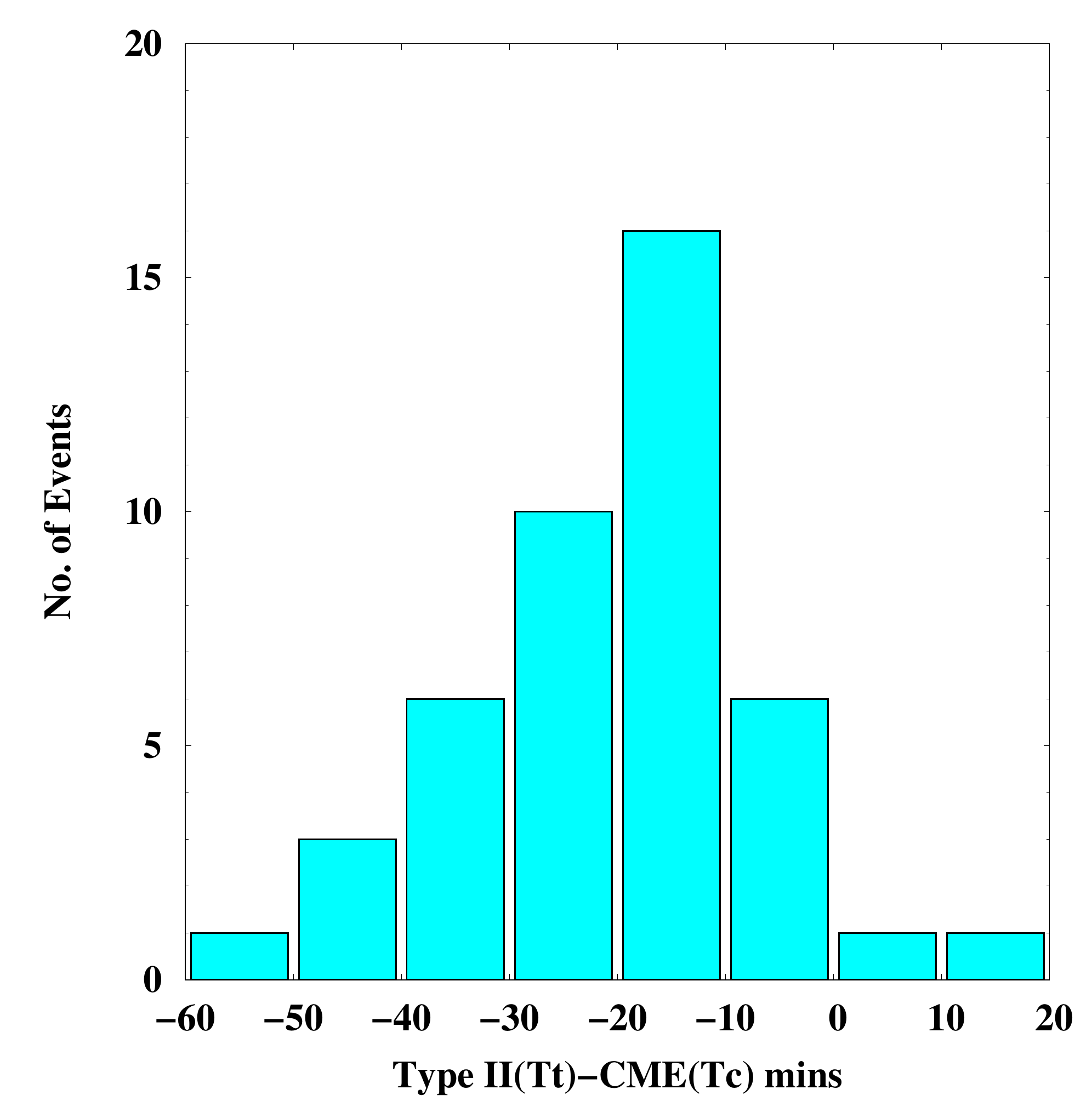}{0.33\textwidth}{(c)}
           }
         
          \caption{Histograms showing time differences of occurence of EUV waves and type II radio bursts (a), EUV waves and CMEs (b) and type II radio bursts and CMEs (c). }
          \label{fig:hist}
\end{figure*}

Figure \ref{fig:hist}(a) shows that out of the 60 events studied, 28 featured the occurrence of type IIs within a time interval of 5 min after the generation of the EUV waves. The maximum delay preceeding type IIs was found to be 25 min after the appearance of the EUV waves. Interestingly, type IIs appeared before EUV waves in five events. A possible physical explanation is that the fast radial motion of the CME generates a shock wave prior to the lateral expansion that produces the EUV wave in the low corona. While these cases can be due to mismatches, their small number is expected to have a minimal effect on our statistical study, which uses a large number of samples. 

Previous works have indicated that CMEs are a common agent of EUV waves and type IIs \citep{Bie02, Mancuso02, Cia05, Gopalswamy05, Gopal09}. In this regard, the properties of CMEs may provide indications as to the link between the EUV waves and type IIs. Figure \ref{fig:hist}(b) and \ref{fig:hist}(c) show the timing difference of the CME onsets from EUV waves and type IIs, respectively. The CME onset times have been taken from the CDAW catalog, and thus the onset time represents the first appearance of CMEs in the LASCO C2 field of view. The larger time delay of CMEs seen in these panels than in Figure \ref{fig:hist}(a) is due to the transit time of the CMEs as they propagate into the C2 field of view. There are few questionable events in which the CME appearance is earlier than that of EUV waves or type IIs, but this should not affect our statistical results.

\begin{figure*}
\centering
\includegraphics[width=1.0\textwidth]{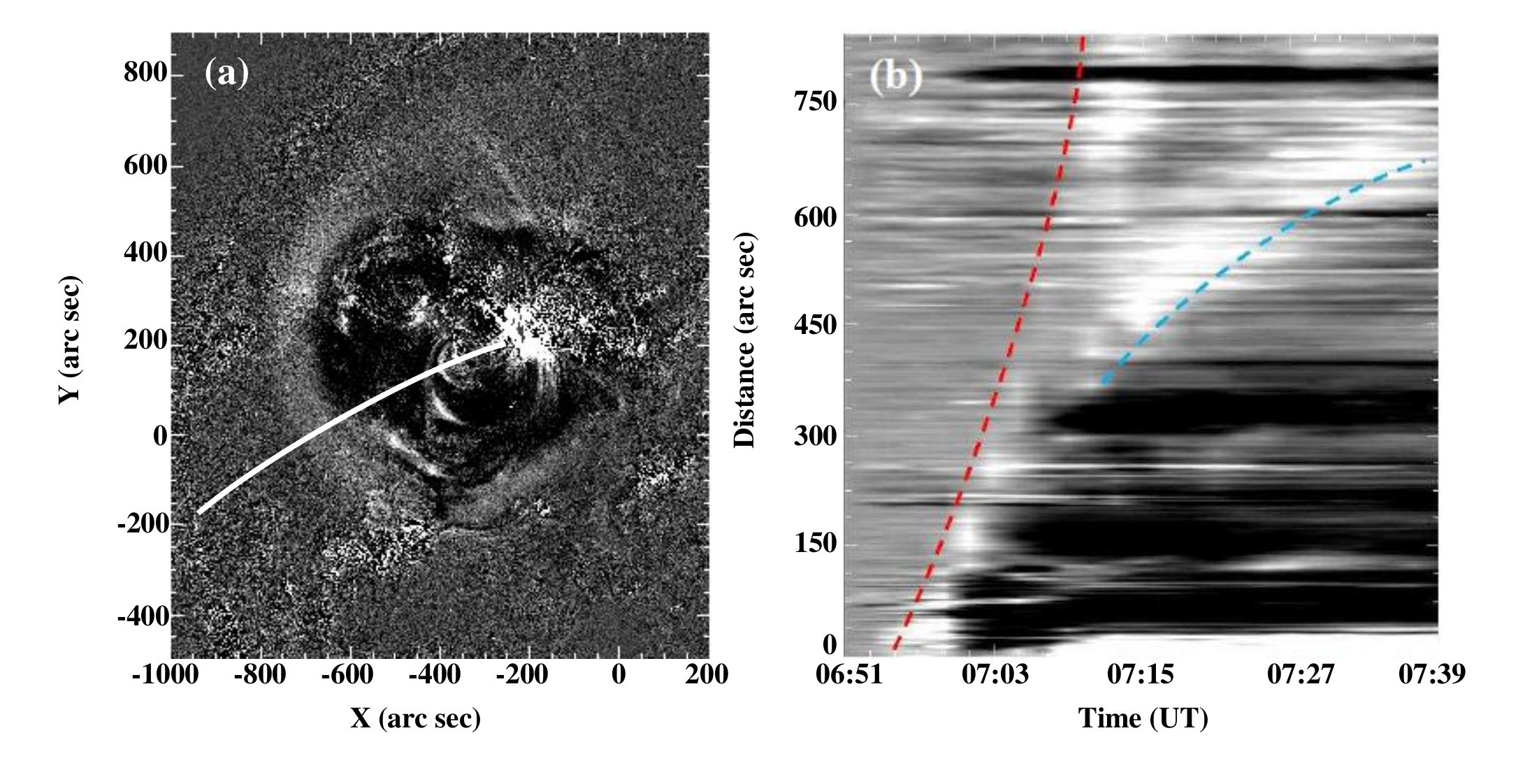}
\caption{(a): Location of the slit along which the distance-time plot is created. (b): Distance-time plot showing the propagating fast-component (red dashed line) and slow-component EUV wave (blue dashed line) on 2013 April 11.} \label{fig:method}

\end{figure*}

We derived the speeds of the 60 EUV waves. Since EUV waves are observed most clearly in 193 \AA\ and 211 \AA\, we selected AIA 193 \AA\ images. Figure \ref{fig:method} shows an example of distance-time maps used to track the EUV wave fronts and determine their speeds. The distance-time map is constructed by stacking cut taken along a great circle passing the source region (the white curve in panel (a) in Figure \ref{fig:method}) on images taken during the EUV wave is in progress. 

To determine the speed, one may use a manual or automatic method \citep[see review by][]{Long2017a}. Note that these methods can be applied only after an EUV wave front is identified by visual inspection. While the different approaches for the same front will result in different speeds, it is very likely that these speeds are correlated each other, and thus these differences may not affect greatly the correlation when comparing with other physical parameters. In this sense, we can speculate that the largest uncertainties in the correlation occur at the selection stage of the fronts due to interference by non-wave components and projection effects (Section \ref{S-Intro}).  

\begin{figure*}
\centering
\includegraphics[width=1.0\textwidth]{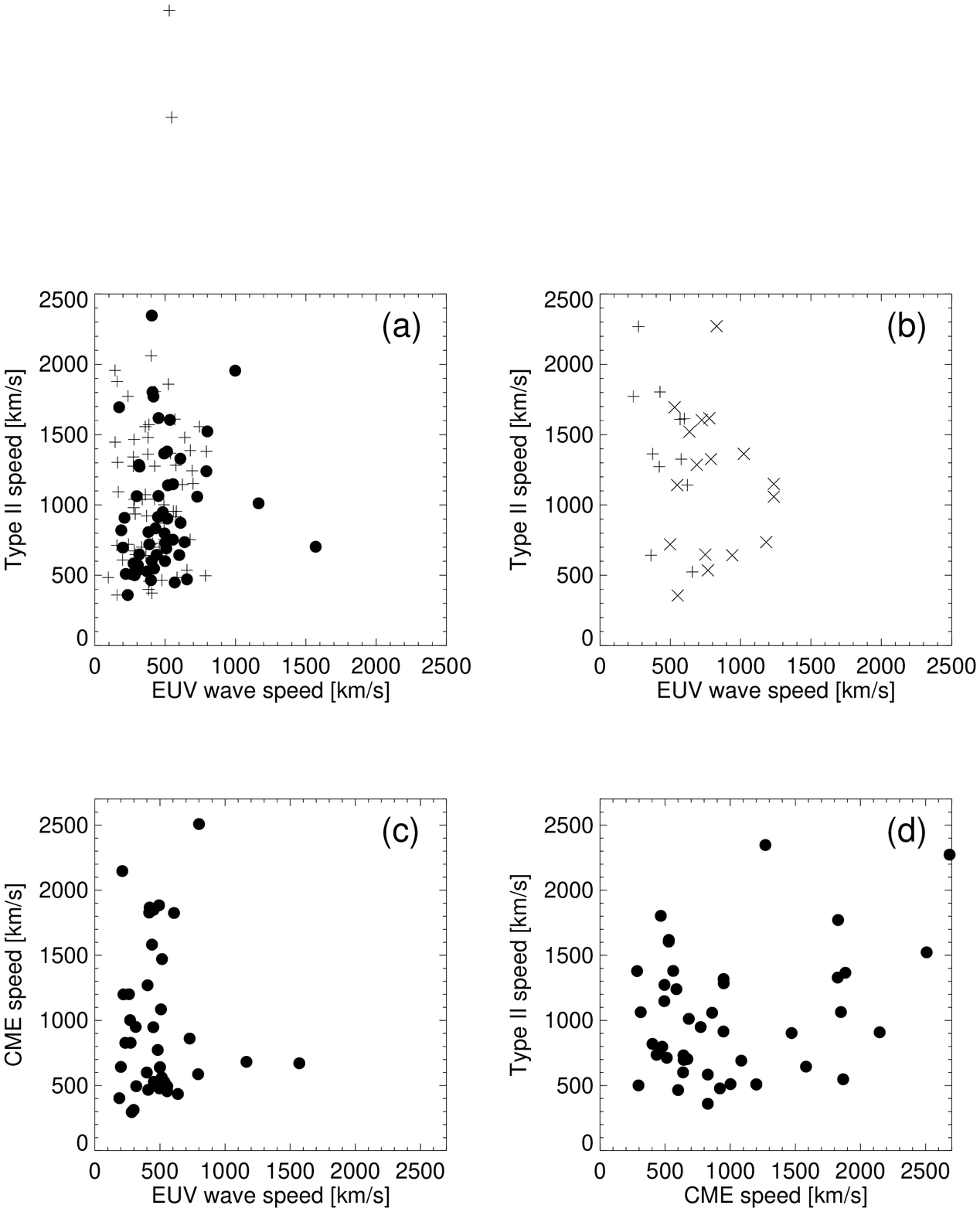}
\caption{Comparisons between speeds of EUV waves, type II radio bursts, and CMEs determined in \citet{Nitta13}, \citet{Long17}, and the present work. See Section \ref{sec:res} for details.} 
\label{fig:comparisons}
\end{figure*}

Taking sources of speed uncertainty into consideration, we used a slightly different method, expecting that comparisons of the given type II speeds with EUV wave speeds determined with various methods may provide indications to find a trend. For every event, the distance-time maps revealed fast as well as slow EUV wave components. We used visual inspection to track these EUV wave-related fronts and computed their average speeds. We found that EUV waves propagate with speeds as low as $\approx$170 km s$^{-1}$ and as high as $\approx$1500 km s$^{-1}$. Note that \citet{Nitta13} and \citet{Long17} used the same catalog and determined the speeds of EUV waves, and thus our investigation can be made together with their speed measurements (Figure \ref{fig:comparisons}).

Figure \ref{fig:comparisons}(a) shows comparisons between the speeds of EUV waves and type IIs. Closed dots denote values determined in this study and show no correlation between the two speeds. Since EUV wave speeds could be uncertain, we also present the values in Figure 6 of \citet{Long17} with cross symbols to determine whether the low correlation is due to the uncertainty of our calculated speeds. Furthermore, the speeds for EUV waves also studied in \citet{Nitta13} and \citet{Long17} are shown in Figure \ref{fig:comparisons}(b) as ex and plus symbols, respectively. These comparisons made with three independent measurements do not provide any indication for a trend. This implies that the inconsistency is not merely due to uncertainties in measuring EUV wave speeds. 
Instead, we found a tendency from these three independent studies, in which the type II speeds easily exceed a higher speed, e.g., 1000 km s$^{-1}$, whereas the EUV waves rarely do so.

The same tendency is also found in the speeds of CMEs and EUV waves, as shown in Figure \ref{fig:comparisons}(c). Their weak correlation was also reported in \citet{Nitta13}. It seems natural that EUV waves are waves propagating with their wave-mode speed while CME speeds depend on kinetic energy. Figure \ref{fig:comparisons}(d) shows that type II speeds are better correlated with those of CMEs (correlation coefficient of 0.29) than with EUV waves (correlation coefficient of 0.17). In addition, the full speed ranges of type IIs and CMEs are identical ($\sim$ 2500 km s$^{-1}$). It also seems natural if type IIs arise due to CME--driven shocks. Note that the CME speeds presented in this paper were determined at the noses. The weak correlation coefficient values between type IIs and CMEs can be understood considering the various source locations of type IIs. \citet{Ramesh12} studied 41 type IIs and found that they could originate from any location of the CME. There are equal possibilities for type II generation from the nose and flank.

\section{Discussion and Conclusion} \label{sec:DC}
We analyzed 60 EUV wave events associated with type IIs. We found no strong correlation between the speeds of EUV waves and type IIs; however, we invoke a tendency that has been regarded as an inconsistency, i.e., that the speeds of EUV waves and their range are much less than those of type IIs (Figure \ref{fig:comparisons}). Type IIs easily exceed higher speeds, e.g., 1000 km\,s$^{-1}$, whereas EUV waves rarely do so. The average and standard deviation (the full range and median value) are 480 $\pm$ 250 km\,s$^{-1}$ (170 -- 1570 km\,s$^{-1}$ and 450 km\,s$^{-1}$), i.e., approximately two times less than those of type IIs, which are 970 $\pm$ 470 km\,s$^{-1}$ (360 -- 2350 km\,s$^{-1}$ and 830 km s$^{-1}$). 

The same tendency has been noted in two other independent studies \citep{kl00,Long17}, suggesting that differences in speed cannot be described only to uncertainties. \citet{Long17} showed that the speed and range of type IIs are three times greater than those of EUV waves. The average speed and standard deviation (the full range and its median value) of EUV waves and type IIs are  430 $\pm$ 170 km\,s$^{-1}$ (100 -- 790 km\,s$^{-1}$ and 410 km\,s$^{-1}$) and 1150 $\pm$ 670 km\,s$^{-1}$ (360 - 4520 km\,s$^{-1}$ and 1040 km\,s$^{-1}$), respectively. According to Figure 4 in \citet{kl00}, the values for EUV waves are 280 $\pm$ 70 km\,s$^{-1}$ (100 -- 500 km\,s$^{-1}$ and 250 km\,s$^{-1}$), whereas those for type IIs are 740 $\pm$ 230 km\,s$^{-1}$ (200 - 1300 km\,s$^{-1}$ and 750 km\,s$^{-1}$). These results also show that the type II speeds and the range are three times greater than those of EUV waves. Meanwhile, we did not see the same tendency between the speeds of the type IIs and CMEs. The average speed and the standard deviation of CMEs are 980 $\pm$ 610 km\,s$^{-1}$, approximately identical to those of the type IIs (970 $\pm$ 470 km\,s$^{-1}$).

This tendency or inconsistency has been expected from the three-dimensional speed distribution of coronal shocks \citep{Kwon17}. \citet{Kwon17} found the same tendency in two shock speeds taken from two different directions in an event, nearly along the radial direction and the direction of the EUV waves. They analyzed three halo CMEs with significantly differing radial speeds in the CDAW catalog. They considered that halo fronts are the observation of coronal shocks \citep{Sheeley00,Kwon15}, and thus the coronal counterparts of EUV waves \citep{Kwon13}. The 3D geometry and kinematics of coronal shocks were determined using the ellipsoid model \citep{Kwon14}. While the shock leading-edge speeds of the three CMEs vary from 1355 to 2157 km\,s$^{-1}$, speeds close to the solar surface, i.e., nearly in the direction of EUV waves, are identical and consistent with the local fast magnetosonic wave speeds (see Figure 16 in \citealp{Kwon17}). It is well-known that type IIs from due to CME-driven shocks propagating with the driver CMEs. Since EUV waves are decaying shock waves decoupled from the driver in the low corona where the driver had left, their speeds should be largely affected by the local fast magnetosonic wave speeds. In this regard, it is not necessary that the speeds of EUV waves and type IIs are correlated.

Considering that EUV waves are a coronal counterpart of Moreton–Ramsey waves, our interpretation above seems contradictory to the strong correlation between the speeds of Moreton–Ramsey waves and type IIs \citep{kl00,Warmuth10}. An explanation for this can be found in the fact that Moreton–Ramsey waves are observed only in very energetic events \citep{Francile16, Cab19, Long19}, and their traveling distances are short compared with those of EUV waves \citep{Warmuth01}. This leads to the conjecture that Moreton-Ramsey waves are strictly related to either the shock driving phase or the initial strong shock phase \citep{Kwon13}, and thus they are a driven shock whose speeds are identical to those of the drivers. On the other hand, EUV waves can be observed even in later phases, when the speeds have reached the local fast magnetosonic wave speed \citep{Warmuth01}. This feature was clearly observed in the simultaneous observation of a Moreton-Ramsey wave and EUV wave in \citet{Asai12}, where they initially propagate cospatially. The Moreton–Ramsey wave disappears first while the EUV wave continues to propagate. Since both Moreton-Ramsey and EUV waves decelerate, the simple average speed over time will result in a slower speed for the one that travels a greater distance \citep{Warmuth01}. In this sense, the speeds of Moreton-Ramsey waves and type IIs, i.e., CME-driven shock, should reflect the driver's radial or expansion speeds, and thus their speeds are very likely to be correlated, as apposed to the cases of EUV waves.

Our interpretation also provides indications for the association problem between EUV waves and type IIs. Type IIs are often observed without EUV waves and {\it vice versa} \citep{Bie02, Thompson09, Muhr14, Nitta13, Long17}. This may depend on the direction of shock wave propagation and whether driven or non-driven shocks are present. Type IIs are observed as a slow frequency drift in the dynamic spectra, and their frequency corresponds to the local electron density. Therefore, shocks strong enough to cause radio emission should travel in the direction of the electron density gradient (nearly radial) and along a sufficiently long distance to be identified as a frequency drift in the dynamic spectra. If a super-magnetosonic CME expansion in the initial phase serves as a piston, but its traveling distance in the direction of the electron density gradient is insufficient, there will be no observable frequency drift. However, once a shock wave is generated in all directions by the 3D piston, a fast magnetosonic wave is decoupled from the piston and can propagate to greater distances as a decaying shock wave, i.e., an EUV waves. 

On the other hand, when piston-driven shocks travel to greater distances, both type IIs and EUV waves will be observed; however, once the EUV wave becomes freely propagating and decouples from the driver, it will be subject to refraction due to the local wave-mode speed gradient \citep{Uchida1968,Wang00,Af11,Kwon13,Kwon17}. Figure 4 in \citet{Kwon17} indicates that the EUV wave cannot be observed due to refraction toward the upper corona in which the local fast magnetosonic speed decreases with height. Additionally, when super-magnetosonic radial motion of a CME occurs with sub-magnetosonic expansion, the shock wave will be driven only in the radial direction. In this case, the EUV wave will not be observed. 

In summary, we have confirmed the inconsistency between the speeds of EUV waves and associated type IIs. The correlation coefficient is 0.17. However, we also identified a tendency for the speeds and range of type IIs to significantly exceed those of the EUV waves. Type IIs can easily exceed 1000 km s$^{-1}$, whereas EUV wave speeds rarely do so. The average and standard deviation of the EUV wave speeds are also significantly less than those of the type IIs and CMEs. This tendency can be also found in \citet{kl00,Warmuth10,Long17}. Despite these differences, the average and standard deviation of type IIs and CMEs are identical and better correlated (correlation coefficient of 0.29). 
These results indicate that type IIs can be as fast as CMEs, as opposed to EUV waves. As indicated by the close relationship between Moreton–Ramsey waves and type IIs \citep{kl00,Warmuth10}, the directionality itself is insufficient to account for this tendency. There must be an additional factor lowering the EUV wave speeds, local fast magnetosonic wave speeds.

We conclude that the inconsistency between the speeds of EUV waves and type IIs is an intrinsic tendency. Type IIs are the consequence of a driven shock in the extended solar corona, and their speeds are thus dependent on those of the driver, i.e., CME. On the other hand, the EUV waves are a non-driven shock wave, propagating in the low corona where the driver had already left; they evolve into a linear wave. The simple average speed of an EUV wave over time could be largely affected by the local fast magnetosonic wave speed and thus be independent of the speeds of both the CME and the accompanied type II. Since the corona Alfv\'en speed and sound speed on average may not vary greatly, the average speed of EUV waves must not show large event-to-event variation relative to type IIs whose speeds are dependent on the speed of the driver CMEs. Our results indicate that this is the case. While our analysis may contain significant uncertainties in identifying EUV waves and type IIs and measuring their speeds, our conclusions lead clearly to a conjecture that the observed inconsistency should remain even if all uncertainties were removed.

\begin{table}

\medskip
\centering
\scalebox{0.77}{
\begin{tabular}{cccccccccccc}
\hline
S. No. & Date & EUV Wave  & GOES  & Flare Start & Wave     & CME Start & CME  & Type II   & Shock  & Type II \\
       &      & Time (UT) & Flare & Time (UT) & Velocity & Time (UT) & Velocity & Time (UT) & Height (R$_{\odot}$) & Velocity\\ 
\hline
1. & 27-01-2012 & 18:08 & X1.7 & 17:37 & 800 & 18:27 & 2508 & 18:10 & 1.15 & 1523 \\
2. & 07-03-2012 & 00:05 & X5.4 & 00:02 &  -- & 00:24 & 2684 & 00:17 & 1.15 & 2273 \\
3. & 07-03-2012 & 01:07 & X1.3 & 01:05 & 608 & 01:30 & 1825 & 01:09 & 1.15 & 1329 \\
4. & 09-03-2012 & 03:39 & M6.3 & 03:22 & 314 & 04:26 & 950 & 03:43 & 1.37 & 1285 \\
5. & 13-03-2012 & 17:15 & M7.9 & 17:12 & 493 & 17:36 & 1884 & 17:15 & 1.15 & 1366 \\
6. & 17-03-2012 & 20:37 & M1.3 & 20:32 & 520 & -- &    --   & 20:38 & 1.15 & 1140 \\
7. & 05-04-2012 & 20:52 & C1.5 & 20:49 & 234 & 21:25 & 828  & 21:08 & 1.68 & 360 \\
8. & 09-04-2012 & 12:13 & C3.9 & 12:12 & -- &  12:36 & 921 &  12:28 & 1.59 & 478 \\
9. & 23-04-2012 & 17:39 & C2.0 & 17:38 & 534 & 18:24 & 528 & 17:42  & 1.42 & 1605 \\
10. & 17-05-2012 & 01:28 & M5.1 & 01:25 & 439 & 01:48 & 1582 & 01:31 & 1.15 & 645 \\
11. & 06-06-2012 & 19:57 & M2.1 & 19:54 & 556 & 20:36 & 494 & 20:03  & 1.21 & 1148 \\
12. & 02-07-2012 & 10:44 & M5.6 & 10:43 & 298 & 11:24 & 313 & 10:47 & 1.27 & 1063 \\
13. & 04-07-2012 & 16:38 & M1.8 & 16:33 & 381 & --    & --  & 16:42 & 1.15 & 807 \\
14. & 06-07-2012 & 23:04 & X1.1 & 23:01 & 417 & 23:24 & 1828 & 23:09 & 1.15 & 1771 \\
15. & 13-08-2012 & 12:35 & C2.8 & 12:33 & 638 & 13:25 & 435 & 12:41 & 1.15 & 736 \\
16. & 21-11-2012 & 06:47 & M1.4 & 06:45 & 387 & --   &  --  & 06:54 & 1.15 & 720 \\
17. & 21-11-2012 & 15:20 & M3.5 & 15:10 & 453 & 16:00 & 529 & 15:33 & 1.15 & 1618 \\
18. & 11-01-2013 & 09:07 & M1.2 & 08:43 & 309 & --  &  --   & 09:14 & 1.17 & 537 \\
19. & 13-01-2013 & 08:35 & M1.7 & 08:35 & 316 & --  & --  &  08:40  & 1.15 & 649 \\  
20. & 18-01-2013 & 17:05 & C5.8 & 16:50 & 173 & --  & --  &  17:10 &  1.15 & 1695 \\
21. & 06-02-2013  & 00:03 & C8.7 & 00:04   & 421 & 00:24 & 1867  & 00:13 &  1.15 & 548          \\
22. & 11-04-2013  & 07:02 & M6.5 & 06:55   & 728  & 07:24 &  861  & 07:02 &  1.18  & 1059 \\
23. & 18-04-2013  & 18:01 & C6.5 & 17:56  &  318  & 18:24 &  495  & 18:23 &  1.57 & 1273 \\




24. & 23-04-2013 & 18:11 & C3.0 & 18:10 & 188 & 18:48 & 403 & 18:23 & 1.49 & 820 \\
25. & 02-05-2013  & 05:02 & M1.1 & 04:58   & 1570  & 05:24 &   671 & 05:06  &  1.15 & 703  \\
26. & 13-05-2013 & 02:00 & X1.7 &  01:53   & 405   & 02:00 & 1270 & 02:10  &  1.15 & 2347 \\
27. & 15-05-2013 & 01:42 & X1.2 & 01:25   & 283   & 01:25     & 296  & 01:37  & 1.15 & 501\\
28. & 30-08-2013 & 02:15 & C8.3 & 02:04   &  --   &  02:48    &  949  & 02:12  & 1.27  & 1318 \\
29. & 11-10-2013  & 07:10 & M1.5 & 07:01   & 220   & 07:24     & 1200 & 07:11 & 1.15 & 510 \\
30. & 13-10-2013  & 00:30 & M1.7 & 00:04   & 496 &  01:25     & 478 & 00:38  & 1.29  & 798 \\
31. & 22-10-2013  & 21:20 & M4.2 & 21:15   & 998  &  --    &  --   & 21:21   & 1.08  & 1955 \\
32. & 25-10-2013  & 07:58 & X1.7 & 07:53   & 793  & 08:12    & 587   & 07:59   & 1.15  & 1240\\
33. & 28-10-2013  & 04:35 & M5.1 & 04:32   & 262 & 04:48      & 1201 & 04:37 & 1.29  & 508 \\
34. & 02-11-2013  & 04:44 & C8.2 & 04:40   & 275 & 04:48     & 828  & 04:46  & 1.15  & 584          \\
35. & 05-11-2013  & 22:10 & X3.3 & 22:07   & 514 & 22:36      & 562  & 22:13   &  0.9  & 1380            \\ 
36. & 08-11-2013  & 04:25 & X1.1 & 04:20   & 431 &  --     & --   & 04:24    & 1.42 & 834                 \\
37. & 10-11-2013  & 05:10 & X1.1 & 05:08   & 1164 & 05:36     & 682   & 05:13    & 1.15  & 1012        \\
38. & 07-12-2013  & 07:18 & M1.2 & 07:17   & 508 & 07:36        & 1085  & 07:27  &  1.15  & 691              \\
39. & 12-12-2013  & 03:15 & C4.6 & 03:11   & 271 & 03:36 & 1002   & 03:17    &  1.54  & 511                \\ 
40. & 07-01-2014  & 18:05 & X1.2 & 18:04   & 452 & 18:24 & 1850  &  18:17   &   1.49  & 1064 \\
41. & 08-01-2014  & 03:46 & M3.6 & 03:39   & 200 & 04:12 &  643  &  03:48   &  1.15   & 697 \\
42. & 11-02-2014  & 03:25 & M1.7 & 03:22   & 610 & --    & --  & 03:33   & 0.9  & 873                \\ 
43. & 20-02-2014  & 07:43 & M3.0 & 07:26   & 449 & 08:00  & 948    & 07:45  & 1.42 & 915     \\
44. & 25-02-2014  & 00:44 & X4.9 & 00:39   & 211 & 01:25 & 2147  & 00:56  &  1.15 & 909       \\
45. & 20-03-2014  & 03:45 & M1.7 & 03:42   & 306 & --  & --   & 03:52   &  1.16 & 572         \\
46. & 28-03-2014  & 19:11 & M2.0 & 19:04   & 372 & --  & --   & 19:18   &  1.26    & 528  \\




47. & 02-04-2014  & 13:18 & M6.5 & 13:18  & 516  & 13:36  & 1471 & 13:23 &  1.28  & 903 \\
48. & 04-04-2014  & 13:43 & C8.3 & 13:34  & 410  & 14:12  & 467  & 13:39 &  1.15  & 1803 \\
49. & 25-04-2014  & 00:20 & X1.3 & 00:17   & 555 & 00:48  & 456  & 00:22  &  1.15 & 753           \\
50. & 08-07-2014  & 16:14 & M6.5 & 16:06   & 483 & 16:36  & 773  & 16:14  &  1.15 & 949 \\
51. & 25-07-2014  & 06:57 & C2.2 & 06:57   & --  & --     & --   & 07:12  &  1.42 & 1090 \\
52. & 22-08-2014  & 10:16 & C2.2 & 10:13  & 399 & 11:12       & 600  & 10:17        &  1.19  & 465         \\
53. & 22-08-2014  &15:43 & C6.2 & 15:40   & 568 & --       & --   &  15:44          &  1.16  & 449       \\
54. & 24-08-2014   & 00:11 & C1.6 & 00:08   &655 & --       & -- & 00:15           & 1.15 & 471 \\
55. & 02-10-2014   & 18:51 & M7.3 & 18:49  & --  & 19:12   &  513  & 18:58 &  1.15  & 713 \\
56. & 30-10-2014   & 13:05 & C9.0 & 13:04  & --  & 13:36   &  285  & 13:08 & 1.15  & 1379 \\
57. & 03-11-2014   & 03:51 & C4.2 & 03:47   & 600 & --      & -- & 03:48            & 1.15 & 644\\
58. & 03-11-2014   & 22:32 & M6.5 & 22:15    & 499 & 23:12      & 638 & 22:33       & 1.15 & 601\\
59. & 06-11-2014   & 03:38 & M5.4 & 03:32    & 501 & 04:00        & 641 &  03:46    & 1.26 & 732\\
60. & 07-11-2014   & 17:04 & X1.6 & 16:53   & 404  & --   & --    &   17:19   &   1.15  & 602 \\

\end{tabular}
}
\vspace*{1cm}
\caption{Timeslice Analysis}

\end{table}

\section*{Acknowledgements}
We thank the referee for his/her constructive comments that helped in improving the manuscript. We acknowledge the SDO team who made AIA data available. SDO is a mission for NASA's Living With a Star (LWS) program. We also thank the online data center NOAA for the type II list and LASCO CDAW Catalog for the list of CMEs. R.-Y.K acknowledges support by the National Research Foundation of Korea (NRF) grant funded by the Korea government (MSIT; project No. 2019-2-850-09).

\bibliographystyle{aasjournal}
\bibliography{references1}

\begin{thebibliography}{}
\expandafter\ifx\csname natexlab\endcsname\relax\def\natexlab#1{#1}\fi
\providecommand{\url}[1]{\href{#1}{#1}}
\providecommand{\dodoi}[1]{doi:~\href{http://doi.org/#1}{\nolinkurl{#1}}}
\providecommand{\doeprint}[1]{\href{http://ascl.net/#1}{\nolinkurl{http://ascl.net/#1}}}
\providecommand{\doarXiv}[1]{\href{https://arxiv.org/abs/#1}{\nolinkurl{https://arxiv.org/abs/#1}}}

\bibitem[{{Afanasyev} \& {Uralov}(2011)}]{Af11}
{Afanasyev}, A.~N., \& {Uralov}, A.~M. 2011, \solphys, 273, 479,
  \dodoi{10.1007/s11207-011-9730-9}

\bibitem[{{Asai} {et~al.}(2012){Asai}, {Ishii}, {Isobe}, {Kitai}, {Ichimoto},
  {UeNo}, {Nagata}, {Morita}, {Nishida}, {Shiota}, {Oi}, {Akioka}, \&
  {Shibata}}]{Asai12}
{Asai}, A., {Ishii}, T.~T., {Isobe}, H., {et~al.} 2012, \apjl, 745, L18,
  \dodoi{10.1088/2041-8205/745/2/L18}

\bibitem[{{Biesecker} {et~al.}(2002){Biesecker}, {Myers}, {Thompson}, {Hammer},
  \& {Vourlidas}}]{Bie02}
{Biesecker}, D.~A., {Myers}, D.~C., {Thompson}, B.~J., {Hammer}, D.~M., \&
  {Vourlidas}, A. 2002, \apj, 569, 1009, \dodoi{10.1086/339402}

\bibitem[{{Brueckner} {et~al.}(1995){Brueckner}, {Howard}, {Koomen},
  {Korendyke}, {Michels}, {Moses}, {Socker}, {Dere}, {Lamy}, {Llebaria},
  {Bout}, {Schwenn}, {Simnett}, {Bedford}, \& {Eyles}}]{Brueckner95}
{Brueckner}, G.~E., {Howard}, R.~A., {Koomen}, M.~J., {et~al.} 1995, \solphys,
  162, 357, \dodoi{10.1007/BF00733434}

\bibitem[{{Cabezas} {et~al.}(2019){Cabezas}, {Asai}, {Ichimoto}, {Sakaue},
  {UeNo}, {Ishitsuka}, \& {Shibata}}]{Cab19}
{Cabezas}, D.~P., {Asai}, A., {Ichimoto}, K., {et~al.} 2019, \apj, 883, 32,
  \dodoi{10.3847/1538-4357/ab3a35}

\bibitem[{{Chen} \& {Wu}(2011)}]{Chen11}
{Chen}, P.~F., \& {Wu}, Y. 2011, \apjl, 732, L20,
  \dodoi{10.1088/2041-8205/732/2/L20}

\bibitem[{{Ciaravella} {et~al.}(2005){Ciaravella}, {Raymond}, {Kahler},
  {Vourlidas}, \& {Li}}]{Cia05}
{Ciaravella}, A., {Raymond}, J.~C., {Kahler}, S.~W., {Vourlidas}, A., \& {Li},
  J. 2005, \apj, 621, 1121, \dodoi{10.1086/427619}

\bibitem[{{Cliver} {et~al.}(1995){Cliver}, {Kahler}, {Neidig}, {Cane},
  {Richardson}, {Kallenrode}, \& {Wibberenz}}]{Cliver95}
{Cliver}, E.~W., {Kahler}, S.~W., {Neidig}, D.~F., {et~al.} 1995, in
  International Cosmic Ray Conference, Vol.~4, International Cosmic Ray
  Conference, 257

\bibitem[{{Cliver} {et~al.}(1999){Cliver}, {Webb}, \& {Howard}}]{Cliver99}
{Cliver}, E.~W., {Webb}, D.~F., \& {Howard}, R.~A. 1999, \solphys, 187, 89,
  \dodoi{10.1023/A:1005115119661}

\bibitem[{{Dal Lago} {et~al.}(2003){Dal Lago}, {Schwenn}, \&
  {Gonzalez}}]{Dal03}
{Dal Lago}, A., {Schwenn}, R., \& {Gonzalez}, W.~D. 2003, Adv. Space Res., 32,
  2637, \dodoi{10.1016/j.asr.2003.03.012}

\bibitem[{{Delaboudini{\`e}re} {et~al.}(1995){Delaboudini{\`e}re}, {Artzner},
  {Brunaud}, {Gabriel}, {Hochedez}, {Millier}, {Song}, {Au}, {Dere}, {Howard},
  {Kreplin}, {Michels}, {Moses}, {Defise}, {Jamar}, {Rochus}, {Chauvineau},
  {Marioge}, {Catura}, {Lemen}, {Shing}, {Stern}, {Gurman}, {Neupert},
  {Maucherat}, {Clette}, {Cugnon}, \& {van Dessel}}]{Del95}
{Delaboudini{\`e}re}, J.-P., {Artzner}, G.~E., {Brunaud}, J., {et~al.} 1995,
  \solphys, 162, 291, \dodoi{10.1007/BF00733432}

\bibitem[{{Domingo} {et~al.}(1995){Domingo}, {Fleck}, \& {Poland}}]{Domingo95}
{Domingo}, V., {Fleck}, B., \& {Poland}, A.~I. 1995, \solphys, 162, 1,
  \dodoi{10.1007/BF00733425}

\bibitem[{{Downs} {et~al.}(2021){Downs}, {Warmuth}, {Long}, {Bloomfield},
  Ryun-Young~{Kwon}, {Vourlidas}, \& {Vr\v{s}nak}}]{Downs21}
{Downs}, C., {Warmuth}, A., {Long}, D.~M., {et~al.} 2021, \apj, 911, 118,
  \dodoi{10.3847/1538-4357/abea78}

\bibitem[{{Francile} {et~al.}(2016){Francile}, {L{\'o}pez}, {Cremades},
  {Mandrini}, {Luoni}, \& {Long}}]{Francile16}
{Francile}, C., {L{\'o}pez}, F.~M., {Cremades}, H., {et~al.} 2016, \solphys,
  291, 3217, \dodoi{10.1007/s11207-016-0978-y}

\bibitem[{{Gopalswamy} {et~al.}(2005){Gopalswamy}, {Aguilar-Rodriguez},
  {Yashiro}, {Nunes}, {Kaiser}, \& {Howard}}]{Gopalswamy05}
{Gopalswamy}, N., {Aguilar-Rodriguez}, E., {Yashiro}, S., {et~al.} 2005,
  Journal of Geophysical Research (Space Physics), 110, A12S07,
  \dodoi{10.1029/2005JA011158}

\bibitem[{{Gopalswamy} {et~al.}(2009){Gopalswamy}, {Thompson}, {Davila},
  {Kaiser}, {Yashiro}, {M{\"a}kel{\"a}}, {Michalek}, {Bougeret}, \&
  {Howard}}]{Gopal09}
{Gopalswamy}, N., {Thompson}, W.~T., {Davila}, J.~M., {et~al.} 2009, \solphys,
  259, 227, \dodoi{10.1007/s11207-009-9382-1}

\bibitem[{{Klassen} {et~al.}(2000){Klassen}, {Aurass}, {Mann}, \&
  {Thompson}}]{kl00}
{Klassen}, A., {Aurass}, H., {Mann}, G., \& {Thompson}, B.~J. 2000, \aaps, 141,
  357, \dodoi{10.1051/aas:2000125}

\bibitem[{{Kwon} {et~al.}(2013){Kwon}, {Ofman}, {Olmedo}, {Kramar}, {Davila},
  {Thompson}, \& {Cho}}]{Kwon13}
{Kwon}, R.-Y., {Ofman}, L., {Olmedo}, O., {et~al.} 2013, \apj, 766, 55,
  \dodoi{10.1088/0004-637X/766/1/55}

\bibitem[{{Kwon} \& {Vourlidas}(2017)}]{Kwon17}
{Kwon}, R.-Y., \& {Vourlidas}, A. 2017, \apj, 836, 246,
  \dodoi{10.3847/1538-4357/aa5b92}

\bibitem[{{Kwon} {et~al.}(2014){Kwon}, {Zhang}, \& {Olmedo}}]{Kwon14}
{Kwon}, R.-Y., {Zhang}, J., \& {Olmedo}, O. 2014, \apj, 794, 148,
  \dodoi{10.1088/0004-637X/794/2/148}

\bibitem[{{Kwon} {et~al.}(2015){Kwon}, {Zhang}, \& {Vourlidas}}]{Kwon15}
{Kwon}, R.-Y., {Zhang}, J., \& {Vourlidas}, A. 2015, \apjl, 799, L29,
  \dodoi{10.1088/2041-8205/799/2/L29}

\bibitem[{{Lario} {et~al.}(2014){Lario}, {Raouafi}, {Kwon}, {Zhang},
  {G{\'o}mez-Herrero}, {Dresing}, \& {Riley}}]{Lario14}
{Lario}, D., {Raouafi}, N.~E., {Kwon}, R.~Y., {et~al.} 2014, \apj, 797, 8,
  \dodoi{10.1088/0004-637X/797/1/8}

\bibitem[{{Lemen} {et~al.}(2012){Lemen}, {Title}, {Akin}, {Boerner}, {Chou},
  {Drake}, {Duncan}, {Edwards}, {Friedlaender}, {Heyman}, {Hurlburt}, {Katz},
  {Kushner}, {Levay}, {Lindgren}, {Mathur}, {McFeaters}, {Mitchell}, {Rehse},
  {Schrijver}, {Springer}, {Stern}, {Tarbell}, {Wuelser}, {Wolfson}, {Yanari},
  {Bookbinder}, {Cheimets}, {Caldwell}, {Deluca}, {Gates}, {Golub}, {Park},
  {Podgorski}, {Bush}, {Scherrer}, {Gummin}, {Smith}, {Auker}, {Jerram},
  {Pool}, {Soufli}, {Windt}, {Beardsley}, {Clapp}, {Lang}, \&
  {Waltham}}]{Lemen12}
{Lemen}, J.~R., {Title}, A.~M., {Akin}, D.~J., {et~al.} 2012, \solphys, 275,
  17, \dodoi{10.1007/s11207-011-9776-8}

\bibitem[{{Long} {et~al.}(2019){Long}, {Jenkins}, \& {Valori}}]{Long19}
{Long}, D.~M., {Jenkins}, J., \& {Valori}, G. 2019, \apj, 882, 90,
  \dodoi{10.3847/1538-4357/ab338d}

\bibitem[{{Long} {et~al.}(2017{\natexlab{a}}){Long}, {Murphy}, {Graham},
  {Carley}, \& {P{\'e}rez-Su{\'a}rez}}]{Long17}
{Long}, D.~M., {Murphy}, P., {Graham}, G., {Carley}, E.~P., \&
  {P{\'e}rez-Su{\'a}rez}, D. 2017{\natexlab{a}}, \solphys, 292, 185,
  \dodoi{10.1007/s11207-017-1206-0}

\bibitem[{{Long} {et~al.}(2017{\natexlab{b}}){Long}, {Bloomfield}, {Chen},
  {Downs}, {Gallagher}, {Kwon}, {Vanninathan}, {Veronig}, {Vourlidas}, {Vr{\v
  s}nak}, {Warmuth}, \& {{\v Z}ic}}]{Long2017a}
{Long}, D.~M., {Bloomfield}, D.~S., {Chen}, P.~F., {et~al.} 2017{\natexlab{b}},
  \solphys, 292, 7, \dodoi{10.1007/s11207-016-1030-y}

\bibitem[{{Mancuso} {et~al.}(2002){Mancuso}, {Raymond}, {Kohl}, {Ko}, {Uzzo},
  \& {Wu}}]{Mancuso02}
{Mancuso}, S., {Raymond}, J.~C., {Kohl}, J., {et~al.} 2002, \aap, 383, 267

\bibitem[{{Moreton}(1960)}]{Moreton60}
{Moreton}, G.~E. 1960, \aj, 65, 494, \dodoi{10.1086/108346}

\bibitem[{{Moreton} \& {Ramsey}(1960)}]{MR60}
{Moreton}, G.~E., \& {Ramsey}, H.~E. 1960, \pasp, 72, 357,
  \dodoi{10.1086/127549}

\bibitem[{{Moses} {et~al.}(1997){Moses}, {Clette}, {Delaboudini{\`e}re},
  {Artzner}, {Bougnet}, {Brunaud}, {Carabetian}, {Gabriel}, {Hochedez},
  {Millier}, {Song}, {Au}, {Dere}, {Howard}, {Kreplin}, {Michels}, {Defise},
  {Jamar}, {Rochus}, {Chauvineau}, {Marioge}, {Catura}, {Lemen}, {Shing},
  {Stern}, {Gurman}, {Neupert}, {Newmark}, {Thompson}, {Maucherat},
  {Portier-Fozzani}, {Berghmans}, {Cugnon}, {van Dessel}, \&
  {Gabryl}}]{Moses97}
{Moses}, D., {Clette}, F., {Delaboudini{\`e}re}, J.-P., {et~al.} 1997,
  \solphys, 175, 571, \dodoi{10.1023/A:1004902913117}

\bibitem[{{Muhr} {et~al.}(2014){Muhr}, {Veronig}, {Kienreich}, {Vr{\v s}nak},
  {Temmer}, \& {Bein}}]{Muhr14}
{Muhr}, N., {Veronig}, A.~M., {Kienreich}, I.~W., {et~al.} 2014, \solphys, 289,
  4563, \dodoi{10.1007/s11207-014-0594-7}

\bibitem[{{Nitta} {et~al.}(2014){Nitta}, {Aschwanden}, {Freeland}, {Lemen},
  {W{\"u}lser}, \& {Zarro}}]{Nitta14}
{Nitta}, N.~V., {Aschwanden}, M.~J., {Freeland}, S.~L., {et~al.} 2014,
  \solphys, 289, 1257

\bibitem[{{Nitta} {et~al.}(2013){Nitta}, {Schrijver}, {Title}, \&
  {Liu}}]{Nitta13}
{Nitta}, N.~V., {Schrijver}, C.~J., {Title}, A.~M., \& {Liu}, W. 2013, \apj,
  776, 58, \dodoi{10.1088/0004-637X/776/1/58}

\bibitem[{{Pesnell} {et~al.}(2012){Pesnell}, {Thompson}, \&
  {Chamberlin}}]{Pesnell12}
{Pesnell}, W.~D., {Thompson}, B.~J., \& {Chamberlin}, P.~C. 2012, \solphys,
  275, 3, \dodoi{10.1007/s11207-011-9841-3}

\bibitem[{{Ramesh} {et~al.}(2012){Ramesh}, {Lakshmi}, {Kathiravan},
  {Gopalswamy}, \& {Umapathy}}]{Ramesh12}
{Ramesh}, R., {Lakshmi}, M.~A., {Kathiravan}, C., {Gopalswamy}, N., \&
  {Umapathy}, S. 2012, \apj, 752, 107, \dodoi{10.1088/0004-637X/752/2/107}

\bibitem[{{Schwenn} {et~al.}(2005){Schwenn}, {dal Lago}, {Huttunen}, \&
  {Gonzalez}}]{Schwenn05}
{Schwenn}, R., {dal Lago}, A., {Huttunen}, E., \& {Gonzalez}, W.~D. 2005,
  Annales Geophysicae, 23, 1033, \dodoi{10.5194/angeo-23-1033-2005}

\bibitem[{{Sheeley} {et~al.}(2000){Sheeley}, {Hakala}, \& {Wang}}]{Sheeley00}
{Sheeley}, N.~R., {Hakala}, W.~N., \& {Wang}, Y.~M. 2000, \jgr, 105, 5081,
  \dodoi{10.1029/1999JA000338}

\bibitem[{{Thompson} \& {Myers}(2009)}]{Thompson09}
{Thompson}, B.~J., \& {Myers}, D.~C. 2009, \apj, 183, 225,
  \dodoi{10.1088/0067-0049/183/2/225}

\bibitem[{{Thompson} {et~al.}(1998){Thompson}, {Plunkett}, {Gurman}, {Newmark},
  {St.~Cyr}, \& {Michels}}]{Thompson98}
{Thompson}, B.~J., {Plunkett}, S.~P., {Gurman}, J.~B., {et~al.} 1998, \grl, 25,
  2465, \dodoi{10.1029/98GL50429}

\bibitem[{{Uchida}(1968)}]{Uchida1968}
{Uchida}, Y. 1968, \solphys, 4, 30

\bibitem[{{Uchida}(1974)}]{Uchida1974}
---. 1974, \solphys, 39, 431

\bibitem[{{Wang}(2000)}]{Wang00}
{Wang}, Y.-M. 2000, \apjl, 543, L89, \dodoi{10.1086/318178}

\bibitem[{{Warmuth}(2010)}]{Warmuth10}
{Warmuth}, A. 2010, Advances in Space Research, 45, 527,
  \dodoi{10.1016/j.asr.2009.08.022}

\bibitem[{{Warmuth}(2015)}]{Warmuth15}
---. 2015, Living Reviews in Solar Physics, 12, \dodoi{10.1007/lrsp-2015-3}

\bibitem[{{Warmuth} {et~al.}(2001){Warmuth}, {Vr{\v s}nak}, {Aurass}, \&
  {Hanslmeier}}]{Warmuth01}
{Warmuth}, A., {Vr{\v s}nak}, B., {Aurass}, H., \& {Hanslmeier}, A. 2001,
  \apjl, 560, L105

\bibitem[{{Wild} \& {McCready}(1950)}]{Wild50}
{Wild}, J.~P., \& {McCready}, L.~L. 1950, Australian Journal of Scientific
  Research A Physical Sciences, 3, 387

\bibitem[{{Yashiro} {et~al.}(2004){Yashiro}, {Gopalswamy}, {Michalek}, {St.
  Cyr}, {Plunkett}, {Rich}, \& {Howard}}]{y04}
{Yashiro}, S., {Gopalswamy}, N., {Michalek}, G., {et~al.} 2004, Journal of
  Geophysical Research (Space Physics), 109, A07105,
  \dodoi{10.1029/2003JA010282}

\end{thebibliography}


\end{document}